\documentclass[english,twocolumn]{revtex4}
\usepackage[T1]{fontenc}
\usepackage[latin9]{inputenc}
\usepackage{amsmath}
\usepackage{esint}
\usepackage{color}
\usepackage{subfigure}

\usepackage{epsfig}
\usepackage{array}
\usepackage{times}
\usepackage{setspace}
\usepackage{mathrsfs}
\usepackage{amssymb}

\usepackage{babel}
\begin{document}
\title{Strongly coupled quantum heat machines}

\author{David Gelbwaser-Klimovsky and Al\'an Aspuru-Guzik}

\affiliation{Department of Chemistry and Chemical Biology, Harvard University,
Cambridge, MA 02138}
\begin{abstract}
Quantum heat machines (QHMs) models generally assume a weak coupling  to the baths.  This supposition is grounded in the separability principle between systems and allows the derivation of the evolution equation for this case. In the weak coupling regime,  the machine's output is limited by the coupling strength, restricting their application. Seeking to overcome this limitation,  we here analyze QHMs in the virtually unexplored strong coupling regime, where separability, as well as other standard thermodynamic assumptions, may no longer hold. We show that strongly coupled QHMs may be as efficient as their weakly coupled counterparts. In addition, we find a novel turnover behavior where  their output saturates and disappears in the limit of ultra-strong coupling.
\end{abstract}
\maketitle

One of the basic tenets of standard thermodynamics is the principle of separability, which allows to clearly define and distinguish systems that interact with each other. When the surface to volume  ratio  is small, 
surface effects are negligible, and thermodynamic variables  only
depend on the volume and not on the shape. This argument implicitly assumes a
weak coupling, restricting the interaction space to a small interface between the systems \cite{CallenBOOK85,Fermibook36,gevajmopt02}.

The assumption of weak coupling 	 was  essential for the development
of open quantum system theory \cite{petruccionebook02}, in particular for the development of the Kossakowski-Lindblad
master equation \cite{davies1commatphy74,gorinijmp76,petruccionebook02}, that   describes the evolution
of a system interacting with a thermal bath.
Quantum heat machines (QHMs) models \cite{levy2012quantum, gelbwaserpre13,quan2007quantum,correa2013performance,zheng2014work,linden2010small,esposito2010quantum,birjukov2008quantum} use this framework to describe the evolution of the ``working fluid'' under   the  influence of the hot and cold baths. Progress in this field has been recently reviewed \cite{kosloff2013quantum,gelbwaserreview15}. QHMs may operate 
 either as  engines, by extracting work power, or as  refrigerators, by investing work power and cooling the cold  bath. In both cases, quantum resources have been proposed \cite{gelbwaser2014heat,rossnagel2014nanoscale,gelbwaser2014power,scully2011quantum} in order to boost their output and efficiency. Nevertheless, these models assume a weak coupling to the baths, resulting in limited  QHMs outputs and consequently restricting their applications.

\begin{figure}[htbp]
	\centering
		\includegraphics[scale=0.9]{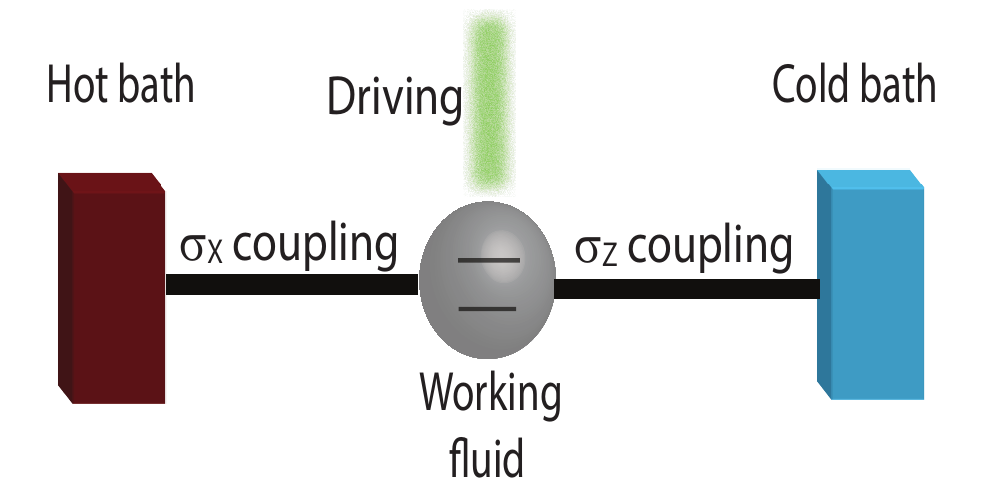}
	\caption{Model of continuous quantum heat machine where the cold bath strongly interacts with the working fluid, while the hot bath is weakly coupled.}
	\label{fig:machine}
\end{figure}

The potential  technological implications of high-output QHMs, such as faster and more powerfull laser cooling \cite{voglnature09,gelbwaser2015laser}, call for a prompt way to overcome the limitation set by the weak coupling assumption. However, the strong coupling limit has been virtually left unexplored due to the lack of theoretical tools to describe the ``working fluid'' evolution. One of the few exceptions \cite{gallego2014thermal} considers the case of   Hamiltonian quench, which involves the switching ``on'' and ``off'' of the system-bath interaction Hamiltonian,  introducing an energy and efficiency cost that reduces the machine efficiency below the Carnot bound. 
 In this letter we take a different approach by  putting forward a strongly coupled continuous QHM model (see Fig. \ref{fig:machine}), that does not require the coupling to and uncoupling from the baths,  which may not be possible at nanoscale, where the system is totally embedded in thermal baths. We investigate its output and efficiency in order to determine its performance limits and which thermodynamic principles, e.g., Carnot bound, still hold at the strong coupling regime. Addressing these issues
becomes more relevant in the light of the large progress achieved
in the field of strongly coupled superconductors \cite{peropadre2013scattering,hoi2011demonstration,hoi2013microwave,astafiev2010resonance}, which makes the
realizations of strongly coupled QHMs potentially tractable in the near future.

\textit{Model and analysis} We employ a model for a continuous QHM similar to the one we studied previously in the weak coupling limit \cite{szczygielskipre13}. This system can operate either as an engine or a refrigerator depending on the spectrum of the reservoirs and the engine's driving frequency. This model is comprised by a driven two-level quantum system,  that represents the working fluid, permanently coupled to the heat baths (hot and  cold).  The evolution of this model is governed by the Hamiltonian

\begin{gather}
\mathcal{H}=\frac{\omega_{0}}{2}\sigma_{z}+\frac{\Omega}{2}(\sigma_{+}e^{-i\omega_{l}t}+\sigma_{-}e^{i\omega_{l}t})+ \nonumber \\
\sigma_{z}\otimes\sum_{k}\xi_{C}(g_{C,k}a_{k}^{\dagger}+g^*_{C,k}a_{k})+ \nonumber \\
\sigma_{x}\otimes\sum_{k}\xi_{H}(g_{H,k}b_{k}^{\dagger}+g^*_{H,k}b_{k})\nonumber \\
+\sum_{k}\omega_{C,k}a_{k}^{\dagger}a_{k}+\sum_{k}\omega_{H,k}b_{k}^{\dagger}b_{k},
\label{eq:hamor}
\end{gather}

\noindent where $\xi_{C(H)}$ is the strength parameter of the cold (hot)
bath,  $g_{i,k}$ is a dimensionless parameter that defines the relative coupling strength of the TLS to the
mode k of the i-bath, $\sigma_{j}$ are the standard Pauli matrices,
and $a_{k}^{\dagger}$, $a_{k}$ ($b_{k}^{\dagger}$,$b_{k}$) are
the creation and annihilation operator of the cold (hot) bath mode $k.$  The election of the hot and cold bath is somehow arbitrary and a similar analysis could be performed if they are interchanged.  

The coupling is consider weak if $\gamma \tau_{cor}\ll 1$, where $\gamma$ is the decay rate and is equivalent to the resonant coupling spectrum ($\gamma=G(\omega_0)$) and $\tau_{cor}$ is the bath correlation time \cite{shahmoon2013nonradiative}. 
Is it  possible to extract work or to cool down in the strong
coupling regime? To elucidate this question, we consider that both couplings are strong. 
While the reduced dynamics is analytically solvable in the weak regime, 
in this case the perturbation expansion on the coupling strength  contains infinite no-neglectable terms  \cite{kryszewski2008master}.

Nevertheless for $\mathcal{H}$, this obstacle may be overcome by solving the problem in a more appropriate basis, where the system is effectively weakly  coupled to the two baths. This is achieved by using  the polaron transformation  \cite{silbey1984variational,parkhill2012correlated,jang2008theory,leejcp12accuracy,devreese1972polarons,mccutcheon2010quantum,chin2011generalized}, $e^S$, where $S=\sigma_{Z}\otimes \sum_{k}(\alpha_{k}a_{k}^{\dagger}-\alpha^*_{k}a_{k})$ and $\alpha_{k}=\xi_{C}\frac{g_{C,k}}{\omega_{C,k}}$. The transformed Hamiltonian,  $\widetilde{\mathcal{H}}=e^{S}\mathcal{H}e^{-s}$, is

\begin{gather}
\widetilde{\mathcal{H}}=\frac{\omega_{0}}{2}\sigma_{z}+\frac{\Omega_{r}}{2}(\sigma_{+}e^{-i\omega_{l}t}+\sigma_{-}e^{i\omega_{l}t})+ \notag\\
\frac{\Omega}{2}\left(e^{-i\omega_{l}t}\sigma_{+}\otimes\left(A_{+}-A\right)+e^{i\omega_{l}t}\sigma_{-}\otimes\left(A_{-}-A\right)\right) +\notag \\
\left(\sigma_{+}\otimes A_{+}+\sigma_{-}\otimes A_{-}\right)\otimes\sum_{k}\xi_{H}(g_{H,k}b_{k}^{\dagger}+g^*_{H,k}b_{k})+ \notag \\
\sum_{k}\omega_{C,k}a_{k}^{\dagger}a_{k}+\sum_{k}\omega_{H,k}b_{k}^{\dagger}b_{k},
\label{eq:Htrans}
\end{gather}
where $A_{\pm}=\Pi_{k}D(\pm2\alpha_{k})$, $D(\alpha_{k})=e^{\alpha_{k}a_{k}^{\dagger}-\alpha^*_{k}a_{k}}$
is the displacement operator, $A=\langle A_{\pm}\rangle=e^{-2\xi_{C}^{2}\sum_{k}\left\Vert \frac{g_{C,k}}{\omega_{k}}\right\Vert ^{2}\coth(\frac{\beta_C\omega_{k}}{2})}$
and $\Omega_{r}=\Omega A$. The terms on the Hamiltonian proportional
to the identity have been neglected.

In the transformed Hamiltonian,  the coupling operators are different. $A_{+}-A$ and $A_{-}-A$, instead of $a_{k}^{\dagger},a_{k}$ and extra terms are added to the hot bath coupling (see Eq. \eqref{eq:Htrans} and Suppl. A). As we show below, the new couplings may be  effectively weak even for high values of the original coupling strengths, $\xi_{C(H)}$. Therefore, the assumptions derived from the weak coupling are correct (e.g. the transformed baths remain at thermal equilibrium) and the master equation may be derived using standard techniques \cite{petruccionebook02,alicki2012periodically} also for values of $\xi_{C(H)}$ that break the weak coupling assumption in the original basis \cite{parkhill2012correlated,jang2008theory,leejcp12accuracy,devreese1972polarons,mccutcheon2010quantum,chin2011generalized}.

The transformed cold bath, now interacts with the TLS through two different operators, $\widetilde{F}_{1}(t)=\frac{\Omega}{2}\left(A_{-}(t)-A\right)$ and $\widetilde{F}_{2}=A_{-}(t)\otimes\sum_{k}\xi_{H}\left(g_{H,k}b_{k}^{\dagger}(t)+g_{H,k}^*b_{k}(t)\right)$.
The  correlation function of the first is

\begin{equation}
\langle\widetilde{F}_{1}(t){}^{\dagger}\widetilde{F_{1}}(0)\rangle=\left(\frac{\Omega A}{2}\right)^{2}(e^{4\xi_{C}^{2}\sum_{k}\frac{\Lambda_{k}(t)}{\omega_{k}^{2}}}-1),
\label{eq:corr1}
\end{equation}
where 

\begin{gather}
\xi_{C}^{2}\sum_{k}\Lambda_{k}(t)=\langle F_{1}^{\dagger}(t)F_{1}(0)\rangle= \notag \\
\sum_{k}\xi_{C}^{2}\left\Vert g_{C,k}\right\Vert ^{2}\left(\cos(\omega_{k}t)\coth(\frac{\beta_C\omega_{k}}{2})-i\sin(\omega_{k}t)\right)
\label{eq:corrweak}
\end{gather}
 is the time correlation of the original coupling operator, $F_1(t)=\sum_{k}\xi_{C}\left(g_{C,k}a_{k}^{\dagger}(t)+g^*_{C,k}a_{k}(t)\right)$ and $\beta_C$ is the equilibrium temperature of the transformed cold bath.
The coupling spectra that govern the evolution  are  derived from the correlations of the transformed operators, $\widetilde{G}_{i}(\omega)=\int_{-\infty}^{\infty}e^{it\omega}\langle\widetilde{F}_{i}(t){}^{\dagger}\widetilde{F}_{i}(0)\rangle dt,\hspace{0.2cm}i\in 1,2$.
 
\begin{figure}
	\includegraphics[scale=0.85]{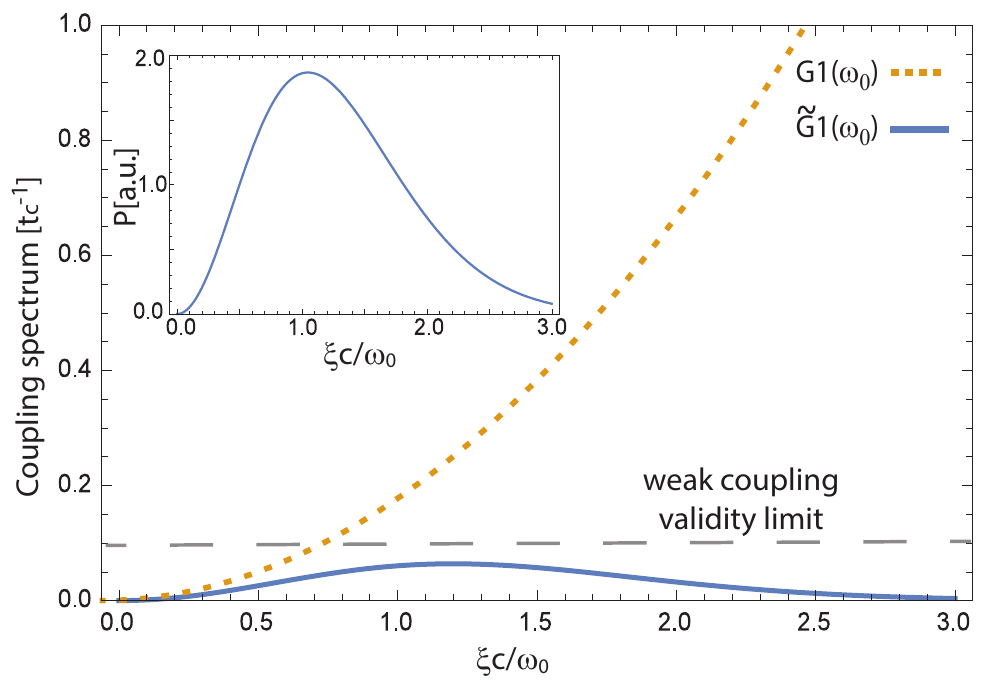}
	\caption{(Color online) Effects of the coupling strength. Main panel: Coupling spectrum in the original basis $G_1(\omega_0)$ (dotted line) and in the transformed basis $\widetilde{G}_1(\omega_0)$ (continuous line)  as a function of the coupling strength $\xi_C\sim \xi_H$. The weak coupling assumption holds only for coupling spectrum below the dashed line. Inset: Power for a QHM as a function of the coupling strength, $\xi_C\sim \xi_H$. A turnover is observed and the power decays for ultra-strong coupling. At this limit,  the lack of   system-bath separability  prevents work extraction.}
	\label{fig:spectrum}
\end{figure}

In Fig. \ref{fig:spectrum},  the dependence on the coupling strength, $\xi_C$, of the  coupling spectrum in the original $\Big($$G_{1}(\omega)=\int_{-\infty}^{\infty}e^{it\omega}\langle F_{1}(t){}^{\dagger}F_{1}(0)\rangle dt$, dotted line$\Big)$ and transformed basis $\Big(\widetilde{G}_1(\omega)$, continuous line$\Big)$ are compared. While both coupling spectra are proportional to the square of the coupling for small coupling strengths, in other regimes their behavior diverge.  The
validity of the ``weak'' coupling assumption for the spectrum $\widetilde{G}_1(\omega)$ has been broadly shown \cite{parkhill2012correlated,jang2008theory,leejcp12accuracy,devreese1972polarons,mccutcheon2010quantum,chin2011generalized,wurger1998strong}.  In a similar manner, the operator $A_{\pm}(t)$, will constraint $\widetilde{G}_2(\omega)$  to the weak coupling regime as long as $\xi_H \sim \xi_C$.

$\widetilde{G}_1(\omega)$
keeps the standard KMS condition $\widetilde{G}_{1}(-\omega)=e^{-\beta_{C}\omega}\widetilde{G}_{1}(\omega)$ \cite{kubo1957statistical,martin1959theory}.
It  includes modes harmonics, i.e., $\widetilde{G}_{1}(\omega>\omega_{cutoff})\neq 0$ as long as  $\omega$ is a linear combination of   bath modes harmonics. This propriety lets the use of \textit{highly detuned baths} in strongly coupled QHMs, unlike for  weakly coupled QHMs that require resonant baths (or at least resonant with linear combinations of $\omega_0$ and $\omega_l$, the TLS  and  driving frequency, respectively \cite{gelbwaserpre13}).

The polaron transformation allows the derivation of the QHM evolution for a wide range of  values of the coupling strengths. Nevertheless, this simplification entails other complications, as  the loss of separability. In the transformed basis, the second correlation  is far from standard. It involves
exchange of excitation with both baths (the operators $b_{k}^{ (\dagger)}$   and  $A_{\pm}$  for the hot and  cold bath respectively).The lack of separability breaks the standard KMS condition, casting doubt on the validity of other thermodynamic principles, as the Carnot bound.

The answer to this question is obtained from the theory of non-equilibrium thermodynamics, which introduces the frequency dependent  ``local'' temperatures, $\beta(\omega)$. They are analogous to the non-equilibrium position-dependent local temperatures \cite{kondepudi2014modern}. In the non-equilibrium framework, the KMS condition is generalized (see Suppl. B):

\begin{gather}
\widetilde{G}_{2}(-\omega)=e^{-\beta(\omega)\omega}\widetilde{G}_{2}(\omega), \notag\\ 
\beta(\omega)= \beta_C \lambda(\omega) +\beta_H \left(1-\lambda \left(\omega \right)\right),
\label{eq:genkms}
\end{gather}
where  $\lambda(\omega)$ measures the relative contribution of the  transformed cold bath to $\widetilde{G}_{2}(\omega)$ and may  take any positive or negative value. Therefore $\beta(\omega)$ is not restricted to the range $[\beta_H,\beta_C]$.  It  depends on both baths coupling strength distribution and modes, making $\beta(\omega)$  frequency dependent, blurring its physical interpretation. As we show later, it  allows to establish clear thermodynamic bounds to the efficiency of the QHMs and to relate them to the Carnot bound. The precise value taken by $\beta(\omega)$, depends on how   the  exchange energy $\omega$ is divided between the hot and the  cold baths.

%
%

In a similar manner as $\widetilde{G}_{1}(\omega)$,  $\widetilde{G}_{2}(\omega)$  not only includes harmonics of the  cold bath, but combinations of them with modes of the hot bath. Therefore, also the hot bath may be highly detuned from the TLS frequency (and from any linear combination with the driving frequency).

In the transformed basis, we  use the standard weak coupling   master equation based on the general Floquet theory of open systems \cite{alicki2012periodically}.  We just stress the main steps, but the detailed derivation may be found in \cite{szczygielskipre13}. The reduced evolution of the TLS density matrix, $\rho$, is given by a linear combination of Lindblad generators obtained from the Fourier components in the interaction picture of the working fluid coupling operators $e^{\mp i\omega_lt}\sigma_{\pm}(t)=\sum_{q\in \mathbf{Z}}\sum_{\omega}S_{1,q}(\omega)e^{-i(\omega+q\omega_l)t}$ and $\sigma_{\pm}(t)=\sum_{q\in \mathbf{Z}}\sum_{\omega}S_{2,q}(\omega)e^{-i(\omega+q\omega_l)t}$.
In the interaction picture,

\begin{gather}
\frac{d \rho}{dt}=  \mathcal{L} \rho, \quad
\mathcal{L}=\sum_{q,\omega} \mathcal{L}^1_{q\omega}+\sum_{q,\omega} \mathcal{L}^2_{q\omega} \notag\\
\mathcal{L}^i_{q\omega}= \frac{G^i(\omega+q\omega_l)}{2} \times \notag\\
 \left(\left[S_{i,q}(\omega)\rho,S_{i,q}^{\dagger}(\omega)\right]+\left[S_{i,q}(\omega),\rho S^{\dagger}_{i,q}(\omega)\right]\right).
\label{eq:evo}
\end{gather}

The TLS  density matrix  evolves until it reaches  a steady state (limit cycle), $\mathcal{L}\bar{\rho}=0$. At this point any transient effect averages out and one may calculate the steady state work power and heat flows, 

\begin{gather}
J_i= \sum_{q \omega} \mathrm{sgn}(\omega)\left(\omega +q\omega_l \right) Tr\left[ \mathcal{L}^i_{q\omega} \bar{\rho}\right], \quad
P=-J_1-J_2,
\label{eq:defth}
\end{gather}
where $\mathrm{sgn}(\omega)=1$ for $\omega>0$ and $\mathrm{sgn}(\omega)=-1$ for $\omega<0$.

In particular we are interested in the ultra-strong coupling regime ($\xi_C\sim\xi_H \gg\omega_0$)  to find out if QHMs may have an ultra-high output. Nevertheless at this limit, and assuming a weak driving with positive detuning ($\delta=\omega_{0}-\omega_{l}\gg \Omega_r>0$), the  work power dependence on the coupling strength goes as (see Suppl. C)

\begin{gather}
P\propto \omega_{l}\frac{A^{2}}{\xi_{C}^{2}}(e^{-\beta_{C}\delta}-e^{-\beta(\omega_{0})\omega_{0}})\propto\frac{e^{-4\xi_{C}^{2}\sum_{k}\left\Vert \frac{g_{C,k}}{\omega_{k}}\right\Vert ^{2}\coth(\frac{\beta_C\omega_{k}}{2})}}{\xi_{C}^{2}}.
\label{eq:power}
\end{gather}

  The conditions for  work extraction  ($P<0$), 
	
\[
\frac{\omega_{l}}{\omega_{0}}<1-\frac{\beta(\omega_0)}{\beta_{C}},
\]
\noindent is derived from Eq. \eqref{eq:power}. 
The heat currents to the baths are:

\begin{gather}
J_1\propto \delta\frac{A^{2}}{\xi_{C}^{2}}(e^{-\beta_{C}\delta}-e^{-\beta(\omega_{0})\omega_{0}})<0, \notag\\
J_2\propto - \omega_0 \frac{A^{2}}{\xi_{C}^{2}}(e^{-\beta_{C}\delta}-e^{-\beta(\omega_{0})\omega_{0}})>0.
\label{eq:currents}
\end{gather}

For  ultra-strong coupled baths, the work power will  decay
with the coupling strength as shown in Eq. \eqref{eq:power}. The exact counterpart of Eq. \eqref{eq:power} for any value of $\xi_C\sim\xi_H$ is plotted on figure \ref{fig:spectrum}-Inset.  Opposite to what may be expected from previous results in the weak coupling regime, work power does not increase indefinitely with the coupling strength. Not only it saturates, but at some point,  decays and vanishes.  At the ultra-strong  limit,  the   system and  the baths  are no longer independent,  preventing work extraction which requires some degree of separability.

The determination of the engine efficiency, as well as the cooling power (in the refrigerator operation mode), is  a more subtle issue. A naive guess would be to consider $J_2$, which is positive, as the incoming heat flow from the hot bath and to define the efficiency as

\begin{equation}
\eta= \frac{-P}{J_2}=\frac{\omega_l}{\omega_0} \leq 1-\frac{\beta(\omega_0)}{\beta_C},
\label{eq:falseeta}
\end{equation}
which can take any value, even above Carnot limit. Nevertheless, the lack of separability  complicates the determination of  how much energy is exchanged with each bath through the coupling spectrum $\widetilde{G}_2(\omega)$. Only a fraction, $1-\lambda(\omega)$, of $J_2$ is originated in the hot bath. Therefore the correct efficiency expression is

\begin{equation}
\eta= \frac{-P}{J_2 \left(1-\lambda\left(\omega_0\right)\right)}=\frac{\omega_l}{\omega_0 \left(1-\lambda(\omega_0)\right)} \leq 1-\frac{\beta_H}{\beta_C}=\eta_{Car}.
\label{eq:eta}
\end{equation}

 From Eq. \eqref{eq:eta} we conclude that \textit{the Carnot bound may be reached}, but not surpassed, by the appropriate choice of driving frequency. Therefore, strongly coupled machines  are as efficient as their weakly coupled counterpart.

 In a similar way, one can calculate the cooling power for the refrigeration operation. This sets the opposite condition on the frequencies: $\frac{\omega_l}{\omega_0} \geq 1-\frac{\beta(\omega_0)}{\beta_C}$, making $J_1>0$, which can erroneously be confused with the cooling power.  The lack of separability
between both baths (the dependence of $\widetilde{F}_{2}$ on both baths operators) mixes the heat flows between both baths and part of $J_2$  is heat flowing to  the cold bath. The correct expression for the cooling power is $J_C=J_1-\lambda(\omega_0) J_2$ and is limited by the Carnot bound for refrigerators. The cooling power has a similar dependence on the coupling strength as the work power and also decays and vanishes for ultra-strong coupling.

An ideal platform to test our results are superconducting quantum circuits, where the almost unexplored strong coupling regime has been recently achieved \cite{niemczyk2010circuit,forn2010observation}, showing astounding $\xi_i /\omega_0$ ratios of around $0.12$.
Moreover, recent theoretical studies have shown that a $\sigma_x-$coupling between quantum microwaves and artificial Josephson-based atoms can be pushed up to  $\xi_i /\omega_0\sim2$ \cite{peropadre2013nonequilibrium}, which is well beyond the critical point ($\xi_i /\omega_0=1$) at which the power efficiency is maximum.
Our proposal consists of a periodically driven superconducting flux qubit with tunable gap \cite{schwarz2013gradiometric}, where the main loop is coupled to the hot bath ($\sigma_x$ coupling), and the $\alpha-$loop is coupled to the cold bath ($\sigma_z$ coupling). In order to bring the $\sigma_z$ coupling to the strong regime, we galvanically couple the $\alpha-$loop to the open transmission line that plays the role of the cold bath.

\textit{Conclusions}
The possibility of work extraction and cooling in the strong coupling regime was shown. Even though some thermodynamic principles, as the standard KMS condition,
do not longer hold at this regime due to the lack of separability between the baths, the operation of the QHMs may be
described  in a non-equilibrium framework. 
This is advantageous because it shows that important principles, as the Carnot bound, still hold in the strong coupling regime.  The introduction of frequency-local
 temperatures, that account for the different baths contributions to the heat flows, are useful to determine how the heat flows are divided between baths and to correctly calculate the QHMs efficiency. As we have shown,  continuous  strongly coupled QHMs, as their weakly coupled counterparts, avoid the efficiency reduction due to the coupling turning on and off and keep the Carnot bound which \textit{can be reached} under the appropriate   driving  frequency.

The appearance of the ``non-equilibrium'' temperatures is related to the loss of separability. Even though both baths, in the transformed basis, are in equilibrium, the heat flows mixes the contribution of both of them, causing an effective deviation from equilibrium. 
 
There are similarities between weakly and strongly coupled QHMs,  but the differences should not be overlooked. While weakly coupled QHMs require baths  with modes resonant to the TLS (or linear combinations with the driving frequency), strongly coupled QHMs  operate also for highly detuned baths, because harmonics of the strongly coupled  bath modes also contribute to both coupling spectra.
An important feature of strongly coupled  QHMs is that,
differently to their weakly coupled counterpart, where the outputs
are proportional to the square of the coupling strength, work and cooling  power
saturate at some point and for ultra-strongly coupled machines they
fall down as the coupling strength increases.  This is a consequence
of the lost of separability as the coupling strength increases, and
shows that QHMs require some degree of separability to operate.

 In order to optimize  QHMs output the ``right'' coupling strength is needed, resembling   the quantum Goldilocks effect \cite{lloyd2011quantum} found in photosynthetic systems.  The latter should be further investigated  to determine if evolution fine-tuned the coupling strength to the baths in order to maximize their chemical power output. Alternatively, the turnover behavior may be corroborated experimentally using superconducting qubits \cite{peropadre2013scattering,hoi2011demonstration,hoi2013microwave,astafiev2010resonance}.

\textit{Acknowledgment} We acknowledge Borja Peropadre Joonssuk Huh for useful discussions. We acknowledge the support from the Center for Excitonics, an Energy
Frontier Research Center funded by the U.S. Department of Energy under award DE-SC0001088. D. G-K. also acknowledges the support of the CONACYT and the  COST Action MP1209.


\newpage
\onecolumngrid

\setcounter{equation}{0}
\renewcommand{\theequation}{S\arabic{equation}}

\section*{Supplementary Information}

\subsection{System-Bath coupling}
In the original basis the system-bath coupling operators are:

\begin{gather}
\sigma_{z}\otimes\sum_{k}\xi_{C}(g_{C,k}a_{k}^{\dagger}+g^*_{C,k}a_{k}), \nonumber \\
\sigma_{x}\otimes\sum_{k}\xi_{H}(g_{H,k}b_{k}^{\dagger}+g^*_{H,k}b_{k}), 
\label{eq:sorint}
\end{gather}
where $\sigma_{i}$ are Pauli matrix and operate on the system. $a_{k}^{\dagger}$ and $a_{k}$ ($b_{k}^{\dagger}$ and $b_{k}$) are the cold (hot) bath operators.

In the transformed basis, the  system-bath coupling operators are

\begin{gather}
\frac{\Omega}{2}\left(e^{-i\omega_{l}t}\sigma_{+}\otimes\left(A_{+}-A\right)+e^{i\omega_{l}t}\sigma_{-}\otimes\left(A_{-}-A\right)\right), \notag \\
\left(\sigma_{+}\otimes A_{+}+\sigma_{-}\otimes A_{-}\right)\otimes\sum_{k}\xi_{H}(g_{H,k}b_{k}^{\dagger}+g^*_{H,k}b_{k}),
\label{eq:stranint}
\end{gather}
where 

\begin{gather}
A_{\pm}-A=\Pi_{k} e^{\pm 2\alpha_{k}a_{k}^{\dagger}\mp 2\alpha^*_{k}a_{k}}-e^{-2\xi_{C}^{2}\sum_{k}\left\Vert \frac{g_{C,k}}{\omega_{k}}\right\Vert ^{2}\coth(\frac{\beta_C\omega_{k}}{2})},  \notag \\
A_{\pm}=\Pi_{k} e^{\pm 2\alpha_{k}a_{k}^{\dagger}\mp 2\alpha^*_{k}a_{k}}.
\label{eq:sop}
\end{gather}


\subsection{Generalized KMS condition}

As mentioned in the main text, the coupling spectrum $\widetilde{G}_{2}(\omega)$
contains	 contributions from both baths. The frequency sum of the contributing
hot and cold bath modes  should match the spectrum frequency,
$\omega=\omega_{H,i}+\omega_{C,j}$. There
are many combinations of modes that match the spectrum frequency, therefore

\begin{equation}
\widetilde{G}_{2}(\omega)=\sum_{i,j}\widetilde{G}_{2}(\omega_{H,i}+\omega_{C,j}).
\label{eq:gsum}
\end{equation}

Due to the non-linearity of the cold bath coupling operators in the transformed basis,  its mode harmonics also contribute to the sum on Eq. \eqref{eq:gsum}. The  $\widetilde{G}_{2}(\omega_{H,i}+\omega_{C,j})$ physical meaning is an energy exchange, where an excitation $\omega$ of the system is interchanged with the $\omega_{H,i}$,
and $\omega_{C,j}$ modes of the hot and cold baths, respectively. They
keep a modified KMS condition

\begin{equation}
\widetilde{G}_{2}(-\omega_{H,i}-\omega_{C,j})=e^{-\beta_{H}\omega_{H,i}-\beta_{C}\omega_{C,j}}\widetilde{G}_{2}(\omega_{H,i}+\omega_{C,j}).
\label{eq:gkms}
\end{equation}

Combining all the terms, the effective frequency-local temperature
may be defined as 

\begin{equation}
e^{-\beta(\omega)\omega}\equiv\frac{\widetilde{G}_{2}(-\omega)}{\widetilde{G}_{2}(\omega)}=\sum_{i,j}e^{-\beta_{H}\omega_{H,i}-\beta_{C}\omega_{C,j}}K_{\omega_{H,i,}\omega_{C},j},\quad K_{\omega_{H,I}\omega_{C,J}}=\frac{\widetilde{G}_{2}(\omega_{H,I}+\omega_{C,J})}{\sum_{i,j}\widetilde{G}_{2}(\omega_{H,i}+\omega_{C,j})},
\label{eq:stemp}
\end{equation}
where  $K_{\omega_{H,I}\omega_{C,J}}$ is the relative weight of the 
$\widetilde{G}_{2}(\omega_{H,I}+\omega_{C,J})$ component.

\subsection{Heat currents and power}
For a weak driving and positive detuning, the heat currents and power are (see \cite{szczygielskipre13}) 

\begin{gather}
J_{1}=\delta\frac{\widetilde{G}_{1}(\delta)\widetilde{G}_{2}(\omega_{0})}{\widetilde{G}_{1}(\delta)+\widetilde{G}_{2}(\omega_{0})}(e^{-\beta_{C}\delta}-e^{-\beta(\omega_{0})\omega_{0}}),\label{eq:spower}\\
J_{2}=-\omega_{0}\frac{\widetilde{G}_{1}(\delta)\widetilde{G}_{2}(\omega_{0})}{\widetilde{G}_{1}(\delta)+\widetilde{G}_{2}(\omega_{0})}(e^{-\beta_{C}\delta}-e^{-\beta(\omega_{0})\omega_{0}}),\\
P=\omega_{l}\frac{\widetilde{G}_{1}(\delta)\widetilde{G}_{2}(\omega_{0})}{\widetilde{G}_{1}(\delta)+\widetilde{G}_{2}(\omega_{0})}(e^{-\beta_{C}\delta}-e^{-\beta(\omega_{0})\omega_{0}}),
\end{gather}
where $\widetilde{G}_{i}(\omega)=\int_{-\infty}^{\infty}e^{it\omega}\langle\widetilde{F}_{i}(t){}^{\dagger}\widetilde{F}_{i}(0)\rangle dt$.

We assume that the main contribution to the coupling spectrum comes
from few modes. For the sake of simplicity we present the calculation assuming
that this contribution is due to one mode. Then,

\begin{gather}
\langle\widetilde{F}_{1}(t){}^{\dagger}\widetilde{F_{1}}(0)\rangle=\left(\frac{\Omega A}{2}\right)^{2}(e^{4\xi_{C}^{2}\sum_{k}\frac{\Lambda_{k}(t)}{\omega_{k}^{2}}}-1)\approx\left(\frac{\Omega A}{2}\right)^{2}(e^{4\xi_{C}^{2}\frac{\Lambda_{k_{0}}(t)}{\omega_{k_{0}}^{2}}}-1) \approx \notag\\
\left(\frac{\Omega A}{2}\right)^{2}\left(\sum_{n=0}^{\infty}2J_{2n}\left(4\frac{\xi_{C}^{2}\left\Vert g_{C,k_{0}}\right\Vert^{2}}{\omega_{k_{0}}^{2}}\right)Cos(2n\omega_{k_{0}}t)+\sum_{n=0}^{\infty}2iJ_{2n+1}\left(4\frac{\xi_{C}^{2}\left\Vert g_{C,k_{0}}\right\Vert^{2}}{\omega_{k_{0}}^{2}}\right)Sin((2n+1)\omega_{k_{0}}t)\right) \times \notag\\
\left(\sum_{n=0}^{\infty}2I_{n}\left(4\frac{\xi_{C}^{2}\left\Vert g_{C,k_{0}}\right\Vert^{2}}{\omega_{k_{0}}^{2}}\coth(\frac{\beta_{C}\omega_{k_{0}}}{2})\right)Cos(n\omega_{k_{0}}t)\right),
\end{gather}
where Bessel and modified Bessel functions have been used to expand
the exponential. Using the Fourier transformation and taking the asymptotic
limits of the Bessel and modified Bessel functions:

\begin{equation}
\widetilde{G}_{1}(\delta)\propto\frac{e^{-4\xi_{C}^{2}\sum_{k}\left\Vert \frac{g_{C,k}}{\omega_{k}}\right\Vert ^{2}\coth(\frac{\beta_{C}\omega_{k}}{2})}}{\xi_{C}^{2}}.
\label{eq:glimit}
\end{equation}

$\widetilde{G}_{2}(\omega_{0})$ has a similar dependence. Therefore,
for $\xi_{C}\rightarrow\infty$, $P\propto\frac{e^{-4\xi_{C}^{2}\sum_{k}\left\Vert \frac{g_{C,k}}{\omega_{k}}\right\Vert ^{2}\coth(\frac{\beta_{C}\omega_{k}}{2})}}{\xi_{C}^{2}}$
.

\end{document}